 \documentclass[prd, showpacs]{revtex4}

\usepackage{dcolumn}
\usepackage{amsmath}

\makeatletter
\def\btt#1{\texttt{\@backslashchar#1}}%
\DeclareRobustCommand\bblash{\btt{\@backslashchar}}%
\makeatother

\begin{document}


\title[Solution to EMD equations]{A self consistent solution to the Einstein Maxwell Dirac Equations }

\author{D. Ranganathan}
 \thanks{Department of Physics, Indian Institute of Technology Delhi, New Delhi 110016 India}

\email{dilip@physics.iitd.ac.in}
\affiliation{
Department of Physics, Indian Institute of Technology Delhi, New Delhi 110016 India}
\date{\today}
\begin{abstract}
A self consistent solution to Dirac equation in a Kerr Newman space-time with $M^2 > a^2 + Q^2$ is presented for the case when the Dirac particle is the source of the curvature and the electromagnetic field. The solution is localised, continuous everywhere and valid only for a special choice of the parameters appearing in the Dirac equation.

\end{abstract}

\pacs{03.65.Ge, 04.20.Jb, 03.65.Pm}

\maketitle

\section{Introduction} \label{Intro:level1}\protect
Bound solutions of the Einstein Maxwell Dirac equations are of interest as a link between quantum mechanics and general relativity in a regime where both theories are well tested. Finster, Smoller and Yau \cite{schwa:1, reis:1, kerr:1}  along with other collaborators showed that no time periodic solutions exist for the Dirac equation in Schwarzschild, Reissner Nordstrom and Kerr Newman ($M^2 > a^2 + Q^2$)  space-times. Dafermos \cite{dafe:1} has confirmed this result by another method. On the other hand Schmid \cite{schmid:1} showed that such time periodic solutions to the Dirac equation do exist for the extreme Kerr case when $M^2 = a^2 + Q^2$.

In these investigations, it was assumed that the Dirac particle mass $m_e$ is very small as compared to the black hole mass $M$, that is, it is an external test particle. I show in the self consistent case, when the Dirac particle is the source of the curvature and the electromagnetic field, that a solution is possible for certain values of the parameters ($m_e, Q, a $) appearing in the Dirac equation. Further this solution is continuous everywhere and normalisable.

Such a solution corresponds to a generalisation of the free particle Dirac equation in Minkowski space to include the effects of the curvature produced by the particle. The ordinary free particle solutions of the Dirac equation are completely delocalised; the curvature however, now causes the Dirac wave functions to be localised over a region comparable in dimension to the Compton wavelength of the particle. Note that the wave function still has an enormous spread in comparison to the dimensions of the event or Cauchy horizons of the particle.

As is usually the case in single particle Dirac theory we ignore the self interaction between the particle and its own electromagnetic field. Further we assume that we need only to look for solutions with zero azimuthal angular momentum about the rotation axis. In the next section the equations resulting from Chandrasekhar's separation of the Dirac equation in a Kerr Newman spacetime \cite{chandra:1, chandra:2} are simplified by applying these assumptions.

A necessary requirement for a wave function is that it be single valued at a point. We carry out a bifurcation analysis of the Dirac equations for $\theta$ and $r$ to find their bifurcation points. The requirement that these lie outside the physical region of interest in both the coordinates $\theta$ and $r$ gives us conditions on the parameters appearing in the Dirac equation. Here we confine ourselves to the simplest case. The equations are then solved and it is found that the angular part of the solution is continuous and normalisable. The radial equation is also solved in each of the three regions (i) the exterior, (ii) the ergosphere between the event and the Cauchy horizons and (iii) the region inside the Cauchy horizon containing the singularity. The resulting radial solutions are continuous in all three regions and so a simple matching of the wave functions at the horizons gives us a unique normalisable solution, with zero energy ($\omega = 0, \; \lambda = 0$).

In the last section, the nature of this solution is discussed. In particular, if the fields are centered at $r = 0$, it is not obvious what it means to have a finite probability of finding the source at $r \neq 0$. I suggest that this can be explained in a manner similar to the explanation for the Klein paradox of Dirac theory. Namely, in the region of the order of the Compton wavelength within which the wave function peaks, the field strengths are so large that virtual pair creation cannot be neglected. The antiparticle thus produced can recombine with the ``original'' particle at $r = 0$ leaving us to detect the particle now at $r \neq 0$. Consider next the case of the only allowed energy eigenvalue zero. For a single bound particle in a potential, the energy can of course be negative or zero.
\section{Solution of the Dirac equation} \label{Solution:level1} \protect

Following the notation of Chandrasekhar's book \cite{chandra:2}, the Kerr Newman line element in Boyer Lindquist coordinates is given by
\begin{subequations}
\label{metric}
\begin{eqnarray}
(ds)^2 & = & \frac{\Delta}{\Sigma} (dt - a {\sin}^2\theta d\phi)^2 - \Sigma (\frac{{d r}^2}{\Delta} + {d \theta}^2 ) - \frac{{\sin}^2 \theta}{\Sigma} ( a d t - (r^2 + a^2 ) d\phi)^2 ,\\
\Sigma & = & r^2 + a^2, \\
\Delta & = & r^2 - 2 M r + a^2 + Q^2 .
\end{eqnarray}
\end{subequations}

As we are using natural units, the mass appearing in the metric $(M \rightarrow \frac{G M}{c^2}) $ has a size comparable to the horizon length. Whereas, the mass $m_e$ which appears in the Dirac equation is $ (m_e \rightarrow \frac{m_e c}{\hbar} )$ has a size which is the inverse of the Compton wavelength. Thus we use different symbols to distinguish the two length scales in the problem. They both of course refer to the same mass but differ according to which of the universal constants $ G, \hbar, c$ was being suppressed.

The Dirac equation can then be separated by writing the four components of the spinor wave function in the form
\begin{subequations}
\label{psi}
\begin{eqnarray}
{\Psi}_1 & = & \frac{R_{-\frac{1}{2}} S_{-\frac{1}{2}}}{r - i a \cos\theta}\: e^{i(m \phi - \omega t)} ,\\ {\Psi}_2 & = &  R_{\frac{1}{2}}   S_{\frac{1}{2}}\: e^{i(m \phi - \omega t)},\\ {\Psi}_3 & = &   R_{\frac{1}{2}} S_{-\frac{1}{2}}\:  e^{i(m \phi - \omega t)},\\ {\Psi}_4 & = &  \frac{R_{-\frac{1}{2}} S_{\frac{1}{2}}}{r + i a \cos\theta}\: e^{i(m \phi - \omega t)}. 
\end{eqnarray}
\end{subequations}

I have slightly modified the notation to use $\omega$ for the frequency rather than $\sigma$. The $R_{\pm\frac{1}{2}}$ are functions of $r$ only while $S_{\pm\frac{1}{2}}$ are functions of $\theta$ only. If this form is substituted into the Dirac equations, \cite{chandra:2} we get the following angular equations.
\begin{subequations}
\label{theta-full}
\begin{eqnarray}
\frac{d}{d \theta} S_{+\frac{1}{2}} & = &-( m \csc\theta + \frac{1}{2} \cot\theta ) S_{+ \frac{1}{2}} - ( \lambda - a m_e \cos\theta) S_{- \frac{1}{2}}, \\
\frac{d}{d \theta} S_{-\frac{1}{2}} & = &( m \csc\theta - \frac{1}{2} \cot\theta ) S_{- \frac{1}{2}} + ( \lambda + a m_e \cos\theta) S_{+ \frac{1}{2}}. \\
\end{eqnarray}
\end{subequations}

We now set $ m = 0$ and carry out a bifurcation analysis of these equations. The angle equations (\ref{theta-full}) will simplify under the substitutions 
\begin{equation}
\label{StoT}
T_{\pm\frac{1}{2}} = \sqrt{\sin\theta} S_{\pm\frac{1}{2}},
\end{equation}
to yield
\begin{subequations}
\label{t-eqns}
\begin{eqnarray}
\frac{d}{d \theta} T_{+\frac{1}{2}} & = & - ( \lambda - a m_e \cos\theta) T_{- \frac{1}{2}}, \\
\frac{d}{d \theta} T_{-\frac{1}{2}} & = & ( \lambda + a m_e \cos\theta) T_{+ \frac{1}{2}}. \\
\end{eqnarray}
\end{subequations}
These have bifurcations at
\begin{equation}
\label{theta-bifur}
(\lambda)^2 = (a m_e \cos\theta)^2.
\end{equation}
Our condition that the wave functions be single valued means that we wish to have no bifurcations in the range $ 0 \leq \theta < \pi $. This is possible if and only if $ |\lambda| \geq a m_e $ or $\lambda = 0$. The second case corresponds to an \textit{s} wave orbital state in the usual separation for the Dirac equation \cite{grein:1} in a central potential. I have only considered this case here.
The angle equations then are
\begin{subequations}
\label{t-s-wave}
\begin{eqnarray}
\frac{d}{d \theta} T_{+\frac{1}{2}} & = &  a m_e \cos\theta T_{- \frac{1}{2}}, \\
\frac{d}{d \theta} T_{-\frac{1}{2}} & = &  a m_e \cos\theta T_{+ \frac{1}{2}}.
\end{eqnarray}
\end{subequations}
 These are trivial to solve and we obtain
\begin{subequations}
\label{-solutions}
\begin{eqnarray}
S_{\frac{1}{2}} & = & A ( e^{a m_e \sin\theta} \pm  e^{- a m_e \sin\theta}) / (\sqrt{2 \sin\theta } ), \\
S_{-\frac{1}{2}} & = & A( e^{a m_e \sin\theta} \mp e^{- a m_e \sin\theta}) / (\sqrt{2 \sin\theta}).
 \end{eqnarray}
\end{subequations}
These solutions while they have a formal singularity at the poles give a finite probability density every where.

Next consider the radial equations. Again I find it convenient to use a slightly different variable from these used by Chandrasekhar (\cite{chandra:2}), we define $ {R^{'}}_{\frac{1}{2}} = \sqrt\Delta R_{\frac{1}{2}}$

\begin{subequations}
\label{rad-full}
\begin{eqnarray}
\frac{d}{d r} {R^{'}}_{+\frac{1}{2}} (r)  & = & i \frac{\omega (r^2 + a^2) + \lambda a}{\Delta} {R^{'}}_{+\frac{1}{2}} (r) + \frac{\lambda - i m_e r}{\sqrt{\Delta}} R_{-\frac{1}{2}} (r), \\
\frac{d}{d r} R_{-\frac{1}{2}} (r) & = &-i \frac{\omega (r^2 + a^2) + \lambda a}{\Delta} R_{-\frac{1}{2}} (r) + \frac{\lambda + i m_e r}{\sqrt{\Delta}} {R^{'}}_{+\frac{1}{2}} (r).
\end{eqnarray}
\end{subequations}
The corresponding bifurcation points lie at those values of $r$ which satisfy
\begin{equation}
\label{rad-birf}
(\omega (r^2 +a^2) +\lambda a )^2 = ({\lambda}^2 + {m_e r}^2 ) \Delta .
\end{equation}
This fourth order equation permits a wide range in the 5 dimensional parameter space $(m_e, \lambda, Q, \omega, m ) $ of the Dirac equation for which the bifurcations lie outside the region of interest. It is worth remarking that such a solution must always have $ {\omega}^2 \geq {m_e}^2 $. Further the solutions with $ \omega = \pm m_e$ are unconditionally stable. As we are considering only the case of no orbital angular momentum, $\lambda = 0$, equation (\ref{rad-birf}) also permits solutions with $\omega = 0$ which have bifurcations at $ r = M \pm \sqrt{\Delta}$ and $r = 0$. I will consider only this last case ($\omega = 0$) in this work.
For $\omega = 0$ and $ \lambda = 0$, the radial equations (\ref{rad-full}) simplify to yield
\begin{subequations}
\label{rad-final}
\begin{eqnarray}
\frac{d}{d r} {R^{'}}_{+\frac{1}{2}} (r)  & = &  \frac{ - i m_e r}{\sqrt{\Delta}} R_{-\frac{1}{2}} (r), \\
\frac{d}{d r} R_{-\frac{1}{2}} (r) & = & \frac{ i m_e r}{\sqrt{\Delta}} {R^{'}}_{+\frac{1}{2}}(r).
\end{eqnarray}
\end{subequations}

Again as in the case of the angle equations, these can be solved quite easily. Consider first the exterior region $ r > M + \sqrt{\Delta} $ if we change the independent variable to $ u = \sqrt{\Delta} + M \ln{( r - M + \sqrt{\Delta})} $ the radial equations (\ref{rad-final}) simplify to yield
\begin{subequations}
\begin{eqnarray}
\frac{d}{d u} {R^{'}}_{+\frac{1}{2}} (u)  & = &  - i m_e R_{-\frac{1}{2}} (u), \\
\frac{d}{d u} R_{-\frac{1}{2}} (u) & = &  i m_e  {R^{'}}_{+\frac{1}{2}} (u).
\end{eqnarray}
\end{subequations}
We can thus write down the solution in the exterior region which is finite as $r \rightarrow \infty $ as
\begin{subequations}
\label{soln-ext}
\begin{eqnarray}
 \sqrt{\Delta} R_{\frac{1}{2}} & = & B \frac{e^{- m_e \sqrt{\Delta}}}{[r - M + \sqrt{\Delta}]^{m_e M}}, \\
R_{-\frac{1}{2}} & = & - i B \frac{e^{- m_e \sqrt{\Delta}}}{[r - M + \sqrt{\Delta}]^{m_e M}}.
\end{eqnarray}
\end{subequations}
Next consider the ergospheric region between the event horizon $ r_+ = M + \sqrt{\Delta} $ and the Cauchy Horizon $ r_{-} = M - \sqrt{\Delta} $. Here we transform the independent variable to $ v = i \sqrt{|\Delta |} +  M \ln{( r - M + i \sqrt{|\Delta |})} $ and obtain the solution
\begin{subequations}
\label{soln-ergo}
\begin{eqnarray}
 \sqrt{\Delta}\! R_{\frac{1}{2}} & = & C e^{i m_e \sqrt{|\Delta |}}[r - M +  i \sqrt{|\Delta |}]^{m_e M} + D \frac{e^{-i m_e \sqrt{|\Delta |}}}{[r - M +  i \sqrt{|\Delta |}]^{m_e M}}, \\
 R_{-\frac{1}{2}} & = & i C e^{i m_e \sqrt{|\Delta |}}[r - M +  i \sqrt{|\Delta |}]^{m_e M} - i D \frac{e^{-i m_e \sqrt{|\Delta |}}}{[r - M +  i \sqrt{|\Delta |}]^{m_e M}}.
\end{eqnarray}
\end{subequations}

Finally in the region interior to the Cauchy horizon $ r \leq r_- $ we use the new independent variable $ w =  \sqrt{\Delta} +  M \ln{( r - M +  \sqrt{\Delta})} $ to obtain the solution
\begin{subequations}
\label{soln-sing}
\begin{eqnarray}
 \sqrt{\Delta}\! R_{\frac{1}{2}} & = & E e^{ m_e \sqrt{\Delta}}[r - M +   \sqrt{\Delta}]^{m_e M} + F \frac{e^{- m_e \sqrt{\Delta}}}{[r - M +   \sqrt{\Delta}]^{m_e M}}, \\
 R_{-\frac{1}{2}} & = & -i E e^{ m_e \sqrt{\Delta}}[r - M +   \sqrt{\Delta}]^{m_e M} + i F \frac{e^{- m_e \sqrt{\Delta}}}{[r - M +   \sqrt{\Delta}]^{m_e M}}.
\end{eqnarray}
\end{subequations}
Thus the solutions are continuous in all three regions, the only effect of the bifurcation at $r_+$ and $r_-$ being an ambiguity as to whether it is my choice in eq.(\ref{soln-ergo} which is correct or its complex conjugate which is to be used in this region. As the solution is continuous in all three regions, the coefficients $ B, C, D, E, F$ can be found up to a normalisation constant by requiring that the solutions be continuous at the horizons. So we finally obtain the solution
\begin{subequations}
\label{soln-final}
\begin{eqnarray}
 \sqrt{\Delta} R_{\frac{1}{2}} & = & B \frac{e^{- m_e \sqrt{\Delta}}}{[r - M + \sqrt{\Delta}]^{m_e M}},\;\;\; ( r \geq r_+ )\\
 \sqrt{\Delta}\! R_{\frac{1}{2}} & = &  B \frac{e^{-i m_e \sqrt{|\Delta |}}}{[r - M +  i \sqrt{|\Delta |}]^{m_e M}}, \;\; ( r_- \leq r \leq r_+ )\\
 \sqrt{\Delta}\! R_{\frac{1}{2}} & = & B e^{ m_e \sqrt{\Delta}}[r - M +   \sqrt{\Delta}]^{m_e M}, \;\; (r \leq r_- ) \\
R_{ - \frac{1}{2}} & = & - i B \frac{e^{- m_e \sqrt{\Delta}}}{[r - M + \sqrt{\Delta}]^{m_e M}}, \;\;\; (r \geq r_+ ) \\
R_{ - \frac{1}{2}} & = &  - i B \frac{e^{-i m_e \sqrt{|\Delta |}}}{[r - M +  i \sqrt{|\Delta |}]^{m_e M}} \;\; ( r_{-} \leq r \leq r_+ ) \\
 R_{ - \frac{1}{2}} & = & -i B e^{ m_e \sqrt{\Delta}}[r - M +   \sqrt{\Delta}]^{m_e M} \;\; (r \leq r_- ).
\end{eqnarray}
\end{subequations}

\section{Discussion} \label{Discussion:level1} \protect

The solution we have just obtained is normalisable and thus represents a proper Dirac spinor wave function. There is of course a second possibility due to the bifurcations at the horizon, which is to take the complex conjugate form inside the ergosphere. This however will give rise to the same Dirac currents etc., outside and is thus indistinguishable form the form presented here.

Next consider the question of the only allowed energy eigenvalue, zero. In any bound state problem, we expect the energy of the state to be less than or equal to the asymptotic value of the binding potential. In this case if we consider the curvature to be producing the attractive force localising the particle close to $ r \approx 0$; then the zero energy eigenvalue should evoke no surprise.

Finally there is the question of what is meant by finding the source of the field at a point other than where the field originates from. Again if this is considered as a potential problem for the Dirac field this can be easily explained. The wave function given by eq.(\ref{soln-final}) is localised over a region of the order of the Compton wavelength of the particle. In this region the field strengths are very high and so virtual pairs of the same particle type have finite amplitude of appearing. The antiparticle of such a pair can then combine with the `` original'' particle leaving the now real particle of the pair to be found in the position $r \neq 0$.

While the present solution might appear overly simplistic, it is the fact that it exists which is important. This suggests that other such solutions may also be found in the rest of the allowed parameter space which was not explored here.

\end{document}